\documentclass[pop,twocolumn,10pt,aip,graphicx,floatfix]{revtex4-1}
\usepackage{graphicx}
\usepackage{color}
\usepackage{float}
\usepackage{subcaption}
\newcommand*{\ten}[1]{\!\times\!10^{#1}}
\newcommand*{\unit}[1]{\;\mathrm{#1}}

\begin{document}

\title{Origins of plateau formation in ion energy spectra under target normal sheath acceleration} 

\author{Timothy C.\ DuBois}
\email[]{timothy.dubois@chalmers.se}
\author{Evangelos Siminos}
\author{Julien Ferri}
\affiliation{Department of Physics, Chalmers University of Technology, Gothenburg, SE-41296 Sweden}
\author{Laurent Gremillet}
\affiliation{CEA, DAM, DIF, F-91297 Arpajon, France}
\author{T\"{u}nde F\"{u}l\"{o}p}
\affiliation{Department of Physics, Chalmers University of Technology, Gothenburg, SE-41296 Sweden}

\date{\today}

\begin{abstract}
    Target normal sheath acceleration (TNSA) is a method employed in laser--matter interaction experiments to accelerate light ions (usually protons).
    Laser setups with durations of a few $10$ fs and relatively low intensity contrasts observe plateau regions in their ion energy spectra when shooting on thin foil targets with thicknesses of order $10 \unit{\mu m}$.
In this paper we identify a mechanism which explains this phenomenon using one dimensional particle-in-cell simulations.
Fast electrons generated from the laser interaction recirculate back and forth through the target, giving rise to time-oscillating charge and current densities at the target backside.
    Periodic decreases in the electron density lead to transient disruptions of the TNSA sheath field: peaks in the ion spectra form as a result, which are then spread in energy from a modified potential driven by further electron recirculation.
The ratio between the laser pulse duration and the recirculation period (dependent on the target thickness, including the portion of the pre-plasma which is denser than the critical density) determines if a plateau forms in the energy spectra.
\end{abstract}

\pacs{}

\maketitle 

\section{Introduction}\label{sec:introduction}
The interaction between an ultra-intense laser pulse and a thin foil target is a promising technique which may offer alternative methods of electron and light-ion acceleration.
The latter offer the possibility to overcome limitations in established systems such as radio-frequency cavity accelerators; for example small acceleration gradients.
Bright, multi-MeV ion beams are a sought after capability for many industries, including, but not limited to medical applications (isotope production, proton radiography, hadron therapy) and energy generation (drivers for fast ignition inertial confinement fusion).
However, without a fundamental understanding of the processes involved in the ultra-short time scales and ultra-high energy regime of the laser-matter interaction, these desired beam properties will remain out of reach.

Whilst advanced acceleration schemes have been proposed (such as hole boring \cite{Schlegel2009}, collisionless shock acceleration (CSA)\cite{Silva2004, Haberberger2012}, light sail \cite{Esirkepov2002, Bulanov2008, Henig2009, Kar2012} and chirped standing wave acceleration \cite{Mackenroth2016}), experimental investigation is currently predominated by the famed target normal sheath acceleration (TNSA) method \cite{Roth2002, Mora2003, Cowan2004, Passoni2010} (for a detailed overview of ion acceleration schemes, see review papers by Tikhonchuk~[\onlinecite{Tikhonchuk2010}], Daido et al.~[\onlinecite{Daido2012}] and Macchi et al.~[\onlinecite{Macchi2013}]).
In contrast to the mechanisms mentioned above which accelerate ions from the frontside (illuminated side) of the target, under TNSA ions are accelerated due to a sheath field formed mainly at the backside due to fast electron expansion (ultra-short, high-contrast pulses are able to generate both front and rear sheaths simultaneously\cite{Ceccotti2007}).

Historically, TNSA has had a number of drawbacks. A few examples are the relatively poor scaling of ion energies with increasing laser intensities\cite{Robson2007} ($\sim I_L^{1/2}$) and poor conversion efficiency in general at high laser powers \cite{Kim2013}.
Even with these impediments, TNSA continues to be a popular experimental method due to its comparatively simple design and implementation in contrast to the advanced schemes listed above.
In addition, many of its deficiencies can be offset by introducing novel laser pulse shapes \cite{Markey2010, Pfotenhauer2010} or target designs \cite{Flippo2008, Buffechoux2010, Burza2011, Gaillard2011, Giuffrida2017}.
TNSA is still the most robust and stable ion accelerating mechanism and continues to improve (for example, maximum proton energies of $85$ MeV with conversion efficiencies of $\sim 7\%\pm 3\%$ have recently been achieved\cite{Wagner2016}).
Therefore, a strong comprehension of the fundamental physics underlying the laser-plasma interaction is of paramount importance to generate functional pulse shapes and targets which satisfy the requirements of perspective applications.

In this paper, we focus on the mechanisms producing modulations in the ion energy spectra, particularly plateau formation.
This phenomenon occurs regularly in experiments and is described in the literature \cite{Clark2000, Kaluza2004, Fukumi2005, Nishiuchi2006, Fang2016, Aurand2016, Senje2017, *SenjeURL}, but is only occasionally investigated in depth.

Early investigations of the TNSA mechanism by Clark et al.~[\onlinecite{Clark2000}] identified proton energy distributions with ``multiple peaks which vary in intensity and position between shots'' and ``two distinct populations with a `flattening' of the spectrum''.
These peaks were suggestively explained by a ``complex electron spectrum, likely to manifest itself as modulations in the ion spectra''.

Subsequent experiments noticed that the high-energy `flattening' or `plateaus' do not form for thin targets (aluminium targets $\lesssim 2\unit{\mu m}$)\cite{Kaluza2004}; $p$-polarisation yields the highest electron conversion efficiency (compared to $s$- and circular-polarisations) -- leading to higher electron energies
and plateau formation\cite{Fukumi2005}; and defocusing of the laser spot removes such a plateau entirely\cite{Aurand2016}.

Two distinct hot and cool electron species have been suggested as the cause of the plateau in the ion spectra\cite{Nishiuchi2006}, where it is assumed that this is due to either natural cooling or the spatial distribution of the electrons.
A strong charge separation theory manages to use this assumption to predict maximum energies of experimental observations, although still dramatically underestimates the number of low-energy protons in the spectrum\cite{Passoni2008}.

Fang et al.~[\onlinecite{Fang2016}] is, to our knowledge, the first investigation to focus solely on this plateau phenomenon, where the existence of an energy plateau is dependent on the laser intensity contrast.
It was found that the higher the contrasts became, the shorter the plateau region, inferring a complex proton generation mechanism at relatively low contrast ($10^{8}$) and strongly suggesting that pre-plasma plays a major part in this mechanism.
Plateaus occurred for a variety of target materials (aluminium and steel) as well as thicknesses, with targets of $6.5\unit{\mu m}$ producing the widest plateau.
For thin targets, it was claimed that a stable acceleration structure may fail to form due to the rapidly forming pre-plasma on the frontside.
Finally, the authors investigate the possibility of a two-process CSA/TNSA mechanism to explain the plateau phenomena.

Recently, experiments at the Lund laser facility investigated the effect of target thickness for aluminium targets $3$, $6$ and $12\unit{\mu m}$ thick (see Ref.\ [\onlinecite{Senje2017}], Figure 3.7, pg 40).
A few notable observations from this experiment were: the number of low-energy protons ($E_p < 2$ MeV) increases with target thickness, while maximum proton energies are lower and both the $6$ and $12 \unit{\mu m}$ targets observe a plateau region in their energy spectrum.

In order to gain better understanding of these experimental and theoretical observations, this paper will investigate the processes that give rise to the plateau phenomenon using particle in cell (PIC) simulations.
We find that a depleted electron density on the backside of the target (due to electron recirculation) causes a temporary disruption in the backside sheath field, disturbing the proton acceleration process.

\section{Simulation Setup}\label{sec:simulation}

The open source PIC code \textsc{EPOCH}\cite{Arber2015} is the tool chosen to simulate the plateau phenomenon.
To achieve highly resolved results using realistic target densities and a wide pre-plasma, the analysis will focus on 1D3P simulations, leaving higher dimensional comparisons for future work.

Our investigation was inspired by the recent experiments at Lund and therefore we will use a simulation setup that resembles the laser profile and aluminium thin foils used in Ref.\ [\onlinecite{Senje2017}].
The Lund laser facility uses a $\lambda_L = 0.8\unit{\mu m}$ wavelength beam at the normalised field strength $a_L = 3.5$ ($I_L \simeq 2.6\ten{19} \unit{W/cm^2}$) and a measured contrast of $\sim\!3\ten{9}$ at $100$ ps before the arrival of the main pulse -- presenting a good estimate of the pre-pulse character.
For these experiments, the main pulse duration was $t_{\mathrm{pulse}} = 38$ fs at full width, half maximum (FWHM) focused to a spot size of $3.5 \unit{\mu m}$ (FWHM).

In our simulations, we construct a simple Gaussian time profile for the laser pulse of duration $t_{\mathrm{pulse}} = 38$ fs, offset by $t_{\mathrm{peak}} = 76$ fs corresponding to the peak intensity of the pulse.
We parametrise time of the simulation below via a $t_0 + N\cdot t_L$ convention, meaning the number $N$ of laser periods $t_L = \lambda_L/c$ after the peak intensity of the laser interacts with the target (here, $t_0 = t_{\mathrm{peak}} + x_0/c$ where $x_0$ denotes the location of the pre-plasma).

Aluminium foil targets $3,\, 6 \text{ and } 12\unit{\mu m}$ thick were tested in the experimental campaign, of which we will investigate only the thickest and thinnest.
The following discussion concerns the $3\unit{\mu m}$ case, unless otherwise stated.
Parameters which are not explicitly mentioned below do not differ for the $12\unit{\mu m}$ target.

The normalised critical density $n_c$ for this laser is $n_c = \omega_L^2 m_e \epsilon_0 / q_e^2 = 1.742\ten{21} \unit{cm^{-3}},\text{ where }\omega_L = 2 \pi c / \lambda_L$; $m_e, q_e$ are the rest mass and charge of an electron respectively and $\epsilon_0$ is the vacuum permittivity.
In terms of this value, we calculate the ion density $n_{Al} = 34.6\,n_c$ for the bulk of the target, corresponding to the mass density of aluminium ($2.7 \unit{g/cm^{-3}}$).
To simplify simulations, we assume that ions will be completely stripped of their electrons via the laser pulse and set the ionisation degree to $Z^* = 13$, then enforce a quasi-neutrality condition on the bulk electron density: $n_{e,\text{bulk}} = n_{Al} Z^* = 449.8 \, n_c$.

A contaminant layer on either side of the target is assumed to exist and is modeled as a thin proton layer, with a thickness of $d_H = 0.02\unit{\mu m}$ and a density $n_H = 3\,n_{Al} = 103.8\,n_c$.
There is no general consensus on the density of these proton layers, since composition, manufacturing, handling and storage all likely play a role in their formation at room temperature (before any laser interaction).
    Studies on a gold substrate identify a value of $n_H \sim 200 n_c$ with a thickness of $10$ \AA\cite{Allen2004}.
This suggests our configuration may be an overestimate to the actual proton density, although these differences should not greatly affect the investigation.
On the frontside, a pre-plasma will form before the main pulse of the laser arrives as a consequence of the ASE pedestal, but simulating this on PIC timescales is not technically feasible.
Instead, we use a hydrodynamic estimation with a flux density of $2\ten{11}\unit{W/cm^2}$ simulated in Ref.\ [\onlinecite{Fang2016}] (Figure 4a), as this is close to the flux density of the ASE pedestal at Lund.
After normalising the coefficients captured from Ref.\ [\onlinecite{Fang2016}] to units of $n_c$ we obtain a pre-plasma electron profile of $n_{e,\text{front}} = 0.0800\exp(0.0763x)+0.0012\exp(0.7150x)$.

\begin{figure}[tb]
    \includegraphics[width=\columnwidth]{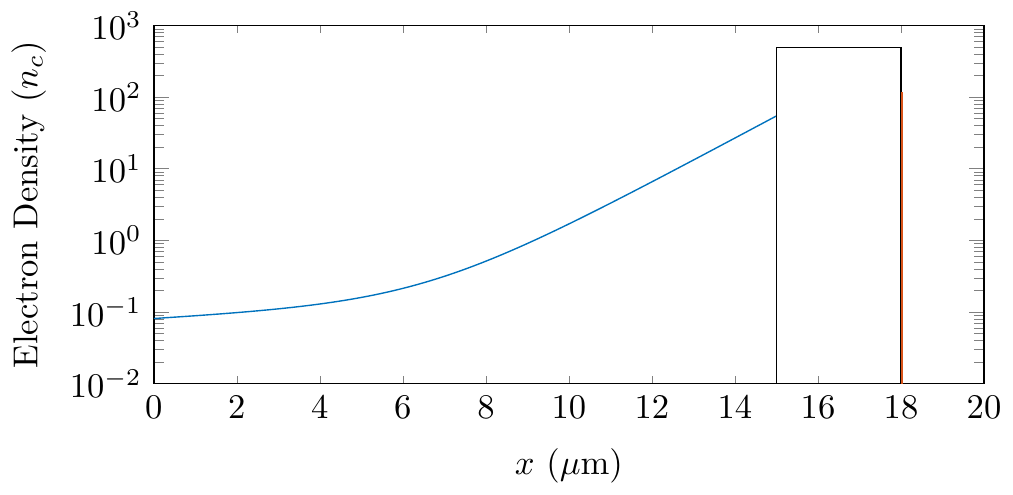}
    \caption{\label{fig:hydrodens} Electron
      density for a $3\unit{\mu m}$ thick aluminum thin foil
      target. Frontside pre-plasma (blue) with a maximum density of
      $54.1 \,n_c$. Bulk (black), $449.8 \,n_c$. Backside proton layer
      (red), $103.8 \,n_c$}
\end{figure}

One more unknown in this instance is the concentration of contaminants such as carbon and oxygen in the pre-plasma, or indeed the ratio of protons to ions originating from the bulk during the pre-plasma expansion process.
Quantification of such values is outside of the scope of this investigation, and for now we construct the pre-plasma entirely out of electrons and protons.
The long rising edge (at very low density, $<0.1\,n_c$) of the pre-plasma is truncated to minimise computational complexity, and from these parameters we build a density profile.
The electrons yield the most complicated structure, which is shown in Figure \ref{fig:hydrodens}.

An estimate of the plasma's initial temperature is calculated from the radiant energy density $w_e = (3/2)n_e k_b T$, where the electron density $n_e = Z^*(T)n_{Al}$ now includes a temperature $T$ dependence.
$w_e$ also depends on $T$ and to solve this self-consistent problem, a Thomas-Fermi approach can be invoked for the ionisation state~\cite{More1985}.
Using the aforementioned flux density and contrast values of the Lund laser we estimate the radiant exposure on target of $H_e = 300\unit{J/cm^2}$ and therefore radiant energy densities of $w_e = 1\ten{6}$ and $2.5\ten{5}\unit{J/cm^3}$ for the $3$ and $12\unit{\mu m}$ targets respectively.
Fitting these values to the Thomas-Fermi model we find initial temperatures $T_0 = 203$ eV for the $3\unit{\mu m}$ target and $T_0 = 52$ eV for the $12\unit{\mu m}$ target.

Electromagnetic boundaries on the simulation domain use the perfectly matched layers algorithm both on the left and right side of the box, with the laser injected on the left side.
The box dimensions are over the range $x = [-50, 50]\unit{\mu m}$.
There are $500$ cells per micron, and the electrons, ions and frontside protons are characterised by $1000$ third order B-spline particles per cell.
The backside proton particle count has been boosted to $750000$ particles per cell to obtain a clean energy spectrum for each run.

\section{Effects of pre-plasma}

Many PIC simulations neglect the effects of pre-plasma or assume some simple exponential profile.
For experiments with a high enough laser contrast, \emph{e.g.} from the inclusion of a double plasma mirror\cite{Levy2007}, neglecting pre-plasma formation is warranted.
Results in the literature\cite{Henig2009a, Dollar2013} that employ such a method, agree that the dominant acceleration scheme is indeed TNSA, yet do not see a plateau in the ion energy profile.
Therefore, it is important to investigate the differences we see between these two target profiles and ascertain if pre-plasma formation is a driving force behind plateau formation.

\begin{figure}[tb]
    \includegraphics[width=\columnwidth]{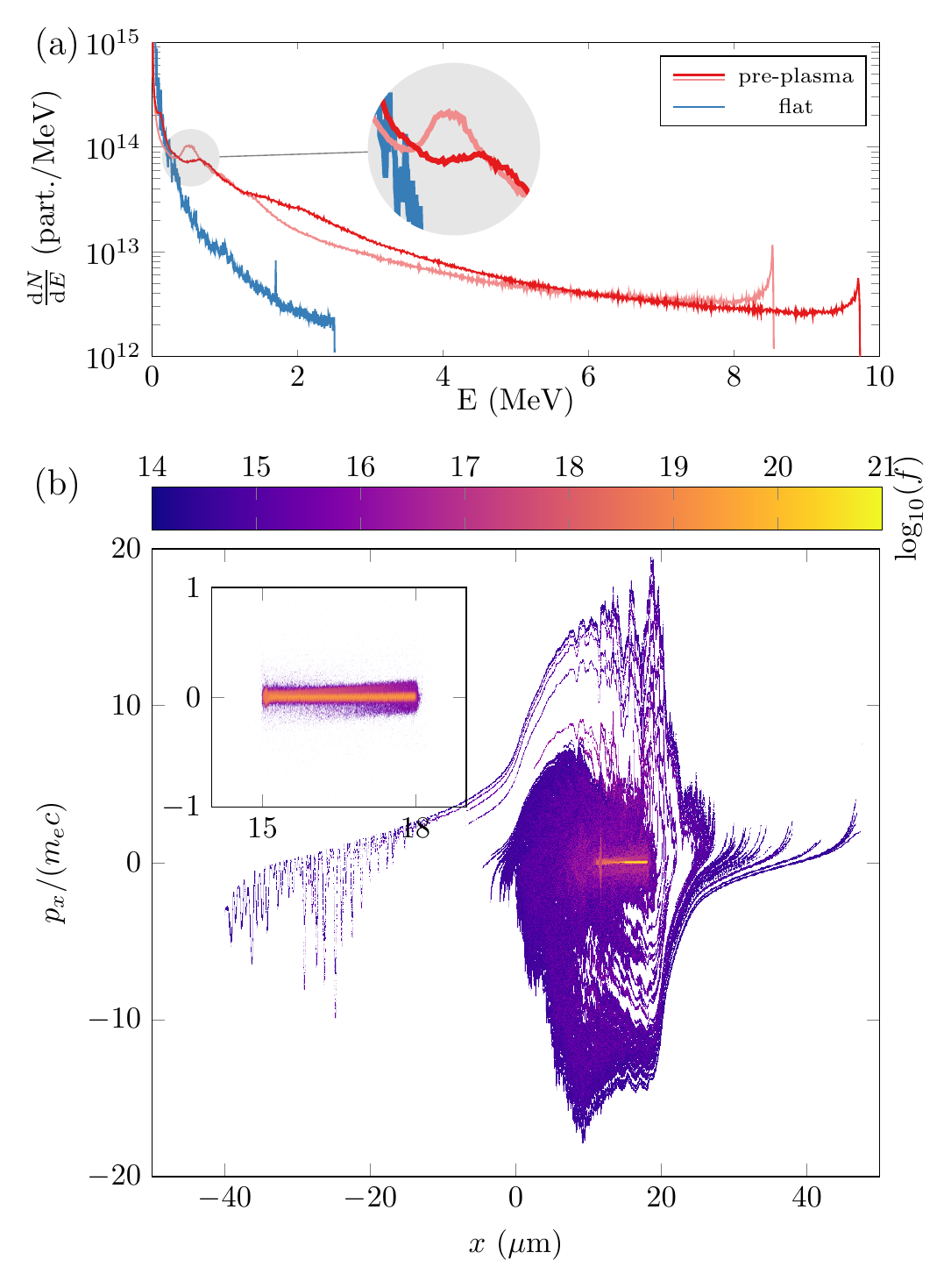}
    \caption{\label{fig:flatcmp} a) Comparison of proton energy
        profiles for $3\unit{\mu m}$ thick targets with (red) and without (blue)
        pre-plasma at isothermal/adiabatic phase boundary $t_{\mathrm{max, iso}}$. A second pre-plasma spectrum (pink) at $t_{\mathrm{max, spot}}$ gives a closer estimate to the maximum proton energy.
        Only protons which have crossed the backside boundary are counted. A peak is observed in the $t_{\mathrm{max, spot}}$ spectrum, which is spread out over time to the extent it is no longer an observable feature on the $t_{\mathrm{max, iso}}$ spectrum. b) Electron
      distributions for the pre-plasma target (main) and flat target
      (inset) at $t = t_0 + 53.7t_L$. Electron heating is almost
      non-existent in the flat target's case, whereas the pre-plasma
      case sees large, anisotropic heating, recirculation and
      clumping.}
\end{figure}

First, we show that the pre-plasma has a large effect on the maximum proton energy.
We note that maximum proton energy for the $3\unit{\mu m}$ target is approximately $8$ MeV in the experimental campaign.
We do not attempt to directly confront the experimental data to our simulations, since, in 1D geometry, the TNSA field is longer-lived than under actual 3D conditions, which tends to overestimate the maximum ion energies.
    For a meaningful comparison, we first consider the time, $t_{\mathrm{max, spot}} = t_0 + 69t_L$, at which the ion front has moved a distance equal to the transverse size of the sheath field (approximately the laser spot size).
A second criterion for setting the maximum simulation time is to identify the isothermal/adiabatic transition, beyond which the ion acceleration process should greatly slow down (while the electron cooling dynamics is sensitive to multidimensional effects)\cite{Mora2005,Brantov2015}.
In the present case, we obtain $t_{\mathrm{max,iso}} = t_0 + 82t_L$.
Backside proton energy spectra for both stopping times are displayed in Figure \ref{fig:flatcmp}a.
Comparing the maximum energy values (at $t = t_{\mathrm{max}}$) for our simulated $3\unit{\mu m}$ targets, we find $E_{\mathrm{max}} = 2.5$ MeV for the flat version and $8.5$ MeV ($t_{\mathrm{max, spot}}$) or $9.7$ MeV ($t_{\mathrm{max, iso}}$) for the target with a pre-plasma.
Due to dimensional effects, we expect 1D simulations to over-estimate energy, so the low value of the flat target clearly illustrates the necessity of including some form of pre-plasma (at the very least a simple exponential) in simulations when comparing to experiments without ultra-high contrasts.
A peak is observed in the $t_{\mathrm{max, spot}}$ spectrum, which can be seen to spread out and dissipate as time evolves to $t_{\mathrm{max, iso}}$.
To identify the mechanism causing this peak, we must extend our simulations outside of this $t_{\mathrm{max}}$ bound, and thus will not discuss direct experimental comparisons here.

In Figure \ref{fig:flatcmp}b we observe complex electron dynamics as a result of strong heating on the front side (compare the minimal heating generated by the flat target with no pre-plasma in the inset).
Recall from Figure \ref{fig:hydrodens} that the laser reflection point is $\sim\!9\unit{\mu m}$ into the pre-plasma -- heating of the large under-dense region occurs both from the initial ponderomotive excitation and in the partially standing wave formed by the interference of the incoming and reflected laser pulse.
The latter effect creates multiple potential wells which the electrons align to as they heat, creating clumping in phase space.
The fast electrons generated in the pre-plasma travel through the target, cross the backside and begin to recirculate.
Recirculation of electrons also occurs on the flat target, albeit with comparatively minuscule velocities.

\begin{figure}[tb]
    \includegraphics[width=\columnwidth]{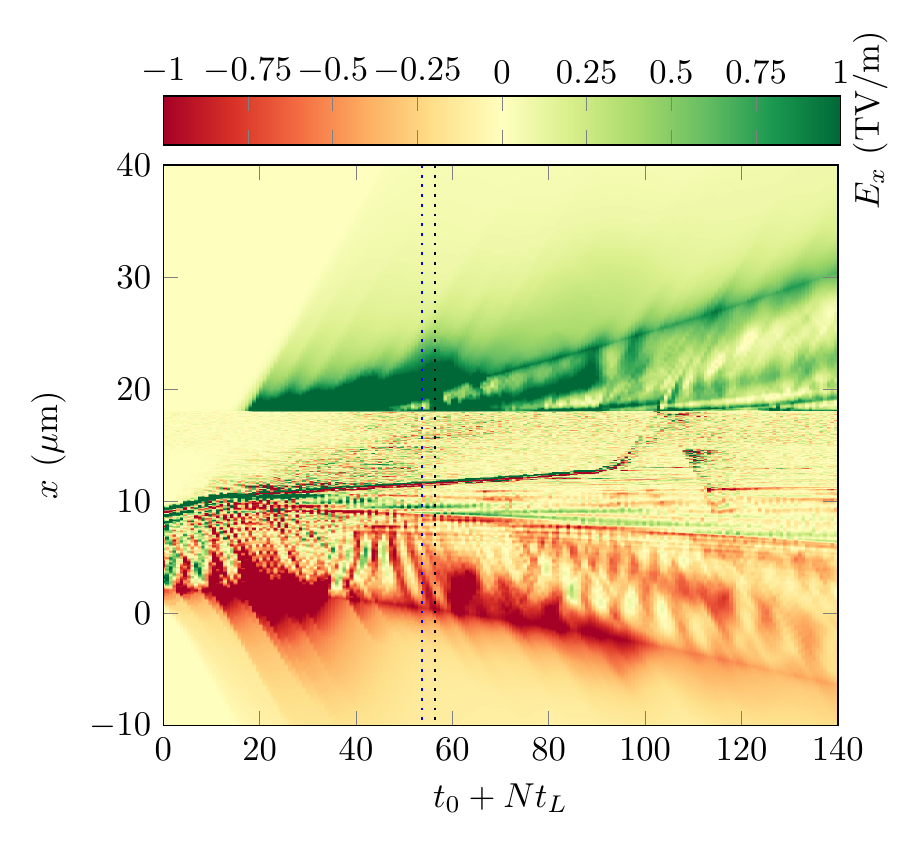}
    \caption{\label{fig:elecfield} Electric field in the $x$ direction
      for a $3\unit{\mu m}$ target with a pre-plasma. On the
      horizontal axis, $N$ represents the axis value of the time formula, where the
      (dashed, blue) line indicates $t = t_0 + 53.7t_L$: the timeslice
      of Figure \ref{fig:flatcmp}b. Notice the sheath depletion begins
      at this time. The maximum field strength is $\sim10$ TV/m, but
      the data range has been truncated here to better visualise the
      field structure. Two regions on the graph are therefore clipped:
      the early TNSA sheath formation region ($15<N<35$) and the CSA
      shock front (the thin, sharp feature running through time at
      around $x=15\unit{\mu m}$). The dashed, black line marks the
      time of the energy spectrum of Figure \ref{fig:energy}.}
\end{figure}

Figure \ref{fig:elecfield} depicts the electric field over the $3 \unit{\mu m}$ target which has two regions of interest relevant to this investigation: the shock front forming in the pre-plasma on the frontside at around $x = 15 \unit{\mu m}$ and the regions of the backside sheath field that seem to have been disrupted.
The streak-like features in the sheath field at early times ($N<60$), which can be seen in Figure \ref{fig:flatcmp}b at $x\gtrsim30$ up to $x\lesssim40\unit{\mu m}$, are due to small populations of fast electrons which escape the target, propagating essentially ballistically at $\sim c$.
The denser streaking occurring later (and at a more acute angle) are a signature of slower ($v\sim 0.16c$), sheath-accelerated ion fronts.
A small depletion region where the field drops from $E_x \sim 5\unit{TV/m}$ down to $\sim0.2\unit{TV/m}$ in the sheath can also be observed here, which we later show to be attributed to electron density fluctuations due to the recirculation process.

\section{Effects of plasma species}\label{sec:species}

Peaks in the proton energy spectrum have been observed previously\cite{Brantov2006}, which manifest via heavy-light ion interactions after light ion acceleration in the sheath field and separation due to an electrostatic shock\cite{Brantov2006, Brantov2009, Diaw2011, Diaw2012}.
These peaks are a feature of the ion spectra at the high energy rather than low energy range, and later we plot an energy spectrum early on in a simulation, observing a high energy peak generated from the ion bulk / backside proton interface due to this mechanism (see Figure \ref{fig:energy}).

\begin{figure}[tb]
    \includegraphics[width=\columnwidth]{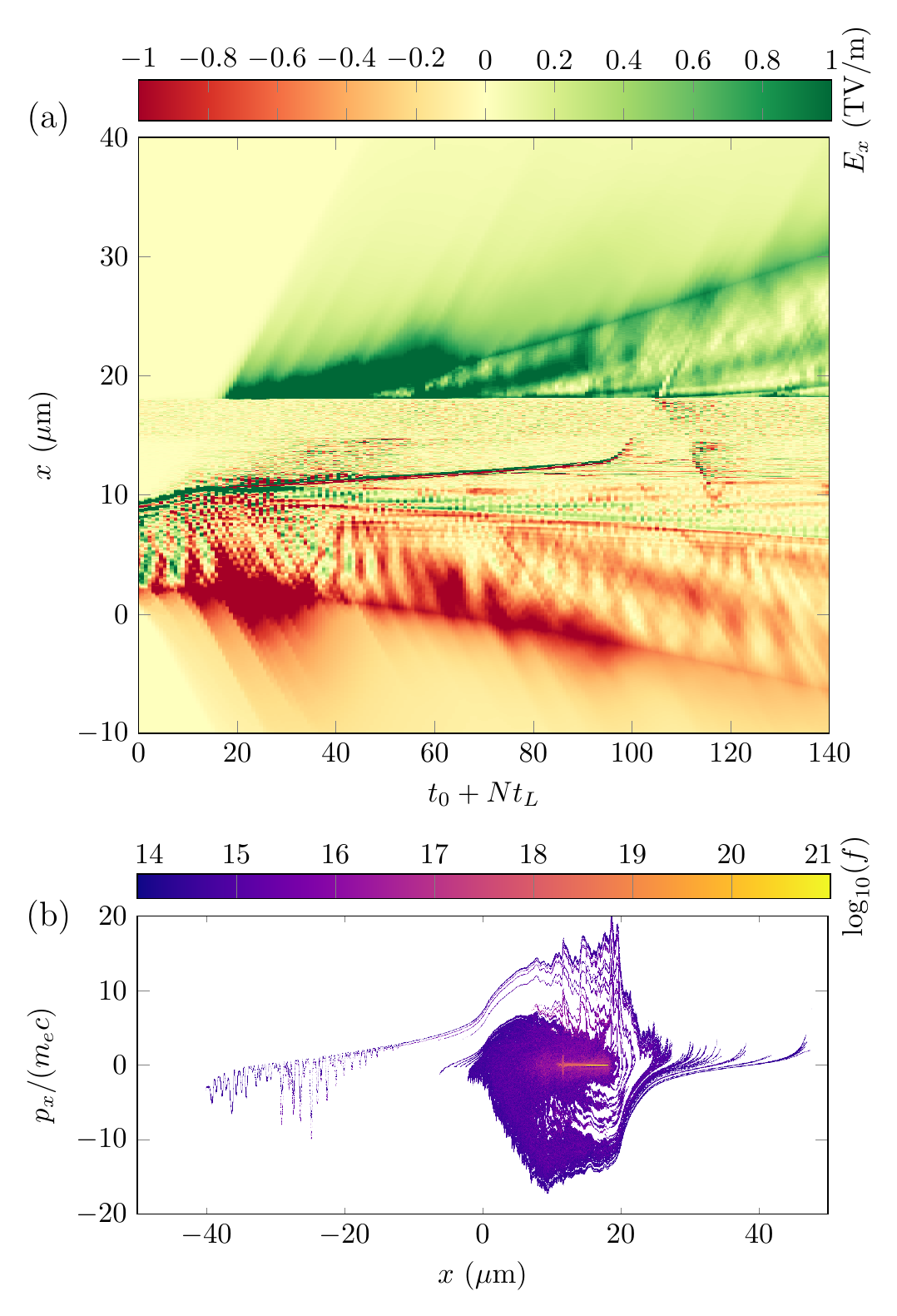}
    \caption{\label{fig:protontarget} a) Electric field in the $x$ direction
        for a theoretical $3\unit{\mu m}$ target, with target and pre-plasma both comprised of just protons and electrons. On the
        horizontal axis, $N$ represents the axis value of the time formula. Notice no major differences when comparing with Figure \ref{fig:elecfield}.
        b) Electron distribution function at $t=t_0+53.7t_L$ to compare directly with Figure \ref{fig:flatcmp}b. Both plots show ion-proton interactions have minimal effect on the interaction dynamics. }
\end{figure}

To investigate definitively whether or not such processes contribute to plateau formation in the current simulations, we generate a non-physical pre-plasma target comprised completely of electrons and protons.
In this case, the ion density in the bulk is set to $n_{i,\,\mathrm{protons}} = 449.8\,n_c$, the same as $n_e$, to keep the quasi-neutrality condition, and all other parameters remain the same.
Figure \ref{fig:protontarget}a depicts the resultant electric field and is mostly consistent with that of Figure \ref{fig:elecfield} (i.e.\ the standard pre-plasma target, showing the same modulations in the expanding proton region).
In addition, the electron distribution (Figure \ref{fig:protontarget}b) remains effectively the same, and the backside proton distribution behaves in the same manner as the standard pre-plasma simulation.
The backside proton energy spectra is altered due to the fact that the bulk is now comprised of protons, which also accelerate in this scenario (bulk ions in the standard simulation are mostly immobile on this time scale).
However, both the energy peaks caused by fast electron activity and peak spreading are still observed.

\section{Effects of target thickness}\label{sec:thickness}

Focusing now on the $12\unit{\mu m}$ target, we investigate the role of target thickness in this scenario.
As discussed in Section \ref{sec:introduction}, decreased maximum energy and the distinct energy plateau are the observed experimental differences.

\begin{figure}[tb]
    \includegraphics[width=\columnwidth]{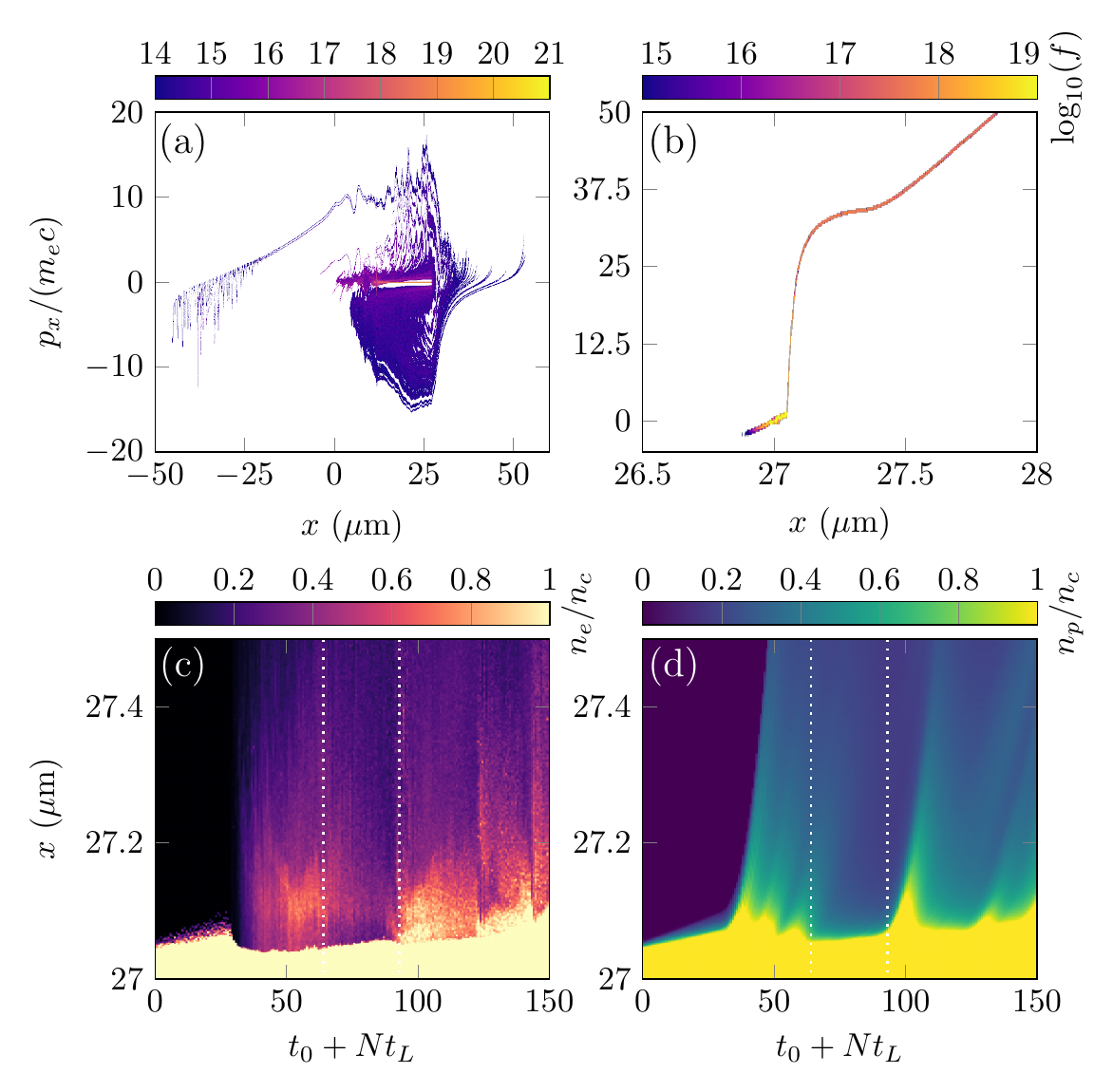}
    \caption{\label{fig:elecdist12} a) Electron distribution of a
      target $12\unit{\mu m}$ thick with a pre-plasma at time
      $t=t_0+61t_L$, representing the time where a visible disruption
      begins in the electric field. b) The resultant proton distribution (focused on the low velocity region) at time $t=t_0+94.9t_L$: after the electric field recovers. The visible kink indicates a modified potential gradient during the disruption event, characterised by the temporal dynamics of the backside electron c) and proton d) densities. Time slices identifying the beginning and end of the disruption (dashed, white) highlight a region of electron depletion as the population recirculates to the frontside of the target, resulting in fewer accelerated protons. The normalised electron and proton densities have been clipped here to differentiate the target and escaping populations.
  }
\end{figure}

The electron distribution of the thick target behaves in a similar manner to the thin one at early times in the simulation: the pre-plasma interaction generates fast electrons which travel through the target to form and later, disrupt, the sheath field.
Figure \ref{fig:elecdist12}a shows the electron distribution function of the thick, $12\unit{\mu m}$ target at the point when the disruption begins.

This distribution can be compared directly to the dynamics of the thin, $3\unit{\mu m}$ target in Figure \ref{fig:flatcmp}b, as they both display the breakdown of the sheath field.
After this point however, the results clearly diverge.

\begin{figure}[tb]
    \includegraphics[width=\columnwidth]{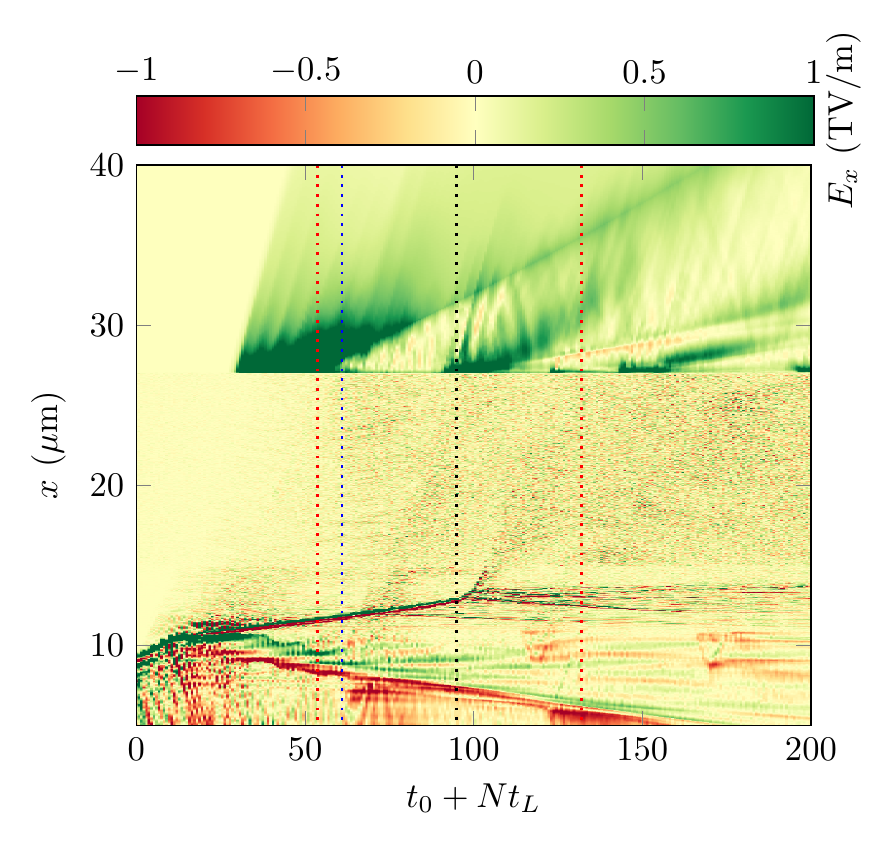}
    \caption{\label{fig:elecfield12} Electric field in the $x$
      direction for a $12\unit{\mu m}$ target. On the horizontal axis,
      $N$ represents the axis value of the time formula, where the
      (dashed, blue) line indicates the $t = t_0 + 61t_L$ time slice
      of Figure \ref{fig:elecdist12}. Dashed, red lines point out time
      slices of the first two energy spectra of Figure
      \ref{fig:ionenergy} and the dashed, black line marks the time of
      the proton distribution of Figure \ref{fig:elecdist12}b and the energy spectrum of Figure \ref{fig:energy}. The maximum
      field strength is $\sim5$ TV/m, but the data range has been
      truncated here to better visualise the field structure. Two
      regions on the graph are therefore clipped: the early TNSA
      sheath formation region ($15<N<35$) and the CSA shock front (the
      thin, sharp feature running through time at around
      $x=15\unit{\mu m}$).}
\end{figure}

Because of the finite duration of the laser pulse, the duration of the fast electron beam is also finite.
After the first electrons begin to recirculate, the electron density eventually decreases significantly in the sheath region, since at some point more electrons travel back towards the target than away from it.
This is depicted in Figure \ref{fig:elecdist12}c, where the electron density begins to diminish close to the backside interface (indicated by the left-most white dashed line at $t=t_0+64t_L$).
The depletion of the electron population leads to a disruption in the electric field of the sheath as seen in Figure \ref{fig:elecfield12}, in which the sheath field on the rear side notably differs from the $3\unit{\mu m}$ target (Figure \ref{fig:elecfield}).
The disruption here is no longer a short pause, but one that lasts for $\sim30t_L$.
We can characterise this time as the period of the fast electron current $t_{\mathrm{sh}}\gtrsim2d/c$ where $d$ is the distance over which the hot electrons recirculate: the target thickness as well as the pre-plasma region where the density is larger that the fast electron density ($n_H \sim n_c$).
For the $12 \unit{\mu m}$ target, the period can be estimated as $t_{\mathrm{sh}}>45t_L\sim 120~$fs: approximately three times that of the pulse duration ($t_{\mathrm{pulse}} = 38~$fs).
This tells us that the head of the fast electron population will still be travelling to the target backside by the time generation of the tail of the population has concluded.
After another half period ($t_{\mathrm{sh}}/2$), most of the electrons are travelling in the $-x$ direction (Figure \ref{fig:elecdist12}a) and the finite length of the population causes the backside density decrease (Figure \ref{fig:elecdist12}c).
As a consequence, we expect a clear disruption of the accelerating field.
Contrarily, the $3 \unit{\mu m}$ target has an estimated current period of $t_{\mathrm{sh}}\sim60~$fs, thus the head of the population has already recirculated and meets the trapping region generated by the laser pulse interaction and as a result, the electron current density at the target backside is hardly modulated.

\begin{figure}[tb]
    \includegraphics[width=\columnwidth]{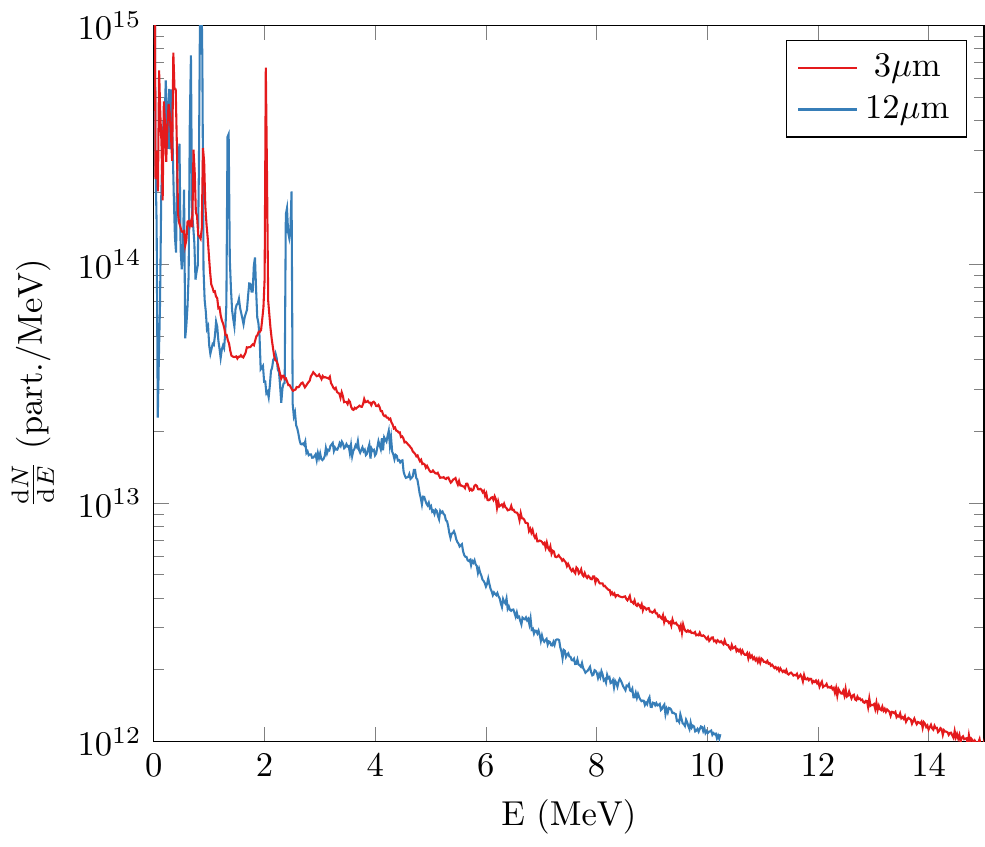}
    \caption{\label{fig:nrgad} Energy spectra of the $3\unit{\mu m}$ (red) and $12\unit{\mu m}$ (blue) targets long into the adiabatic phase ($t=t_0+500t_L$). A plateau has formed between $3$ and $5$ MeV in the $12\unit{\mu m}$ case, which is not observed in the $3\unit{\mu m}$ spectra. }
\end{figure}

For the proton population, the electron depletion means a decrease in the accelerating field and therefore fewer accelerated protons.
We see this effect between the dashed, white markers on the proton density (Figure \ref{fig:elecdist12}d), and most notably in the proton distribution after the sheath recovers from the disruption (Figure \ref{fig:elecfield12}).
Figure \ref{fig:nrgad} compares the energy spectra of the $3\unit{\mu m}$ and $12\unit{\mu m}$ targets long into the adiabatic phase at $t=t_0+500t_L$.
    The highest energy peaks (around $2$ MeV) in both spectra are due to backside proton acceleration caused by an electron jet coming from the breakdown of the frontside shock as it interacts with the target bulk.
    More importantly, a plateau region exists between $3$ and $5$ MeV for the $12\unit{\mu m}$ case, while no such phenomenon is observed in the $3 \unit{\mu m}$ case (for reasons that will be explored in Section \ref{sec:plateau}).

Fang et al.~[\onlinecite{Fang2016}] concentrate their analysis of the plateau formation mechanism on two time slices of their data: $120t_L$ and $200t_L$ and suggest that the CSA protons are the major driver in the formation of an energy plateau.
Frontside protons are shown to be accelerated by a shock moving through the bulk at $120t_L$ reaching the backside at $200t_L$.
At this point in time these CSA protons had not been accelerated via the TNSA process.
However, they conclude that a plateau in the energy spectrum is the result of CSA protons -- stemming from a plot of the spectra at $120t_L$.

Our previous work\cite{SvedungWettervik2016} investigated the CSA/TNSA interplay with highly resolved Vlasov-Maxwell simulations.
The system parameters of that investigation are not widely dissimilar to the ones in Ref.~[\onlinecite{Fang2016}].
It was confirmed that the number density of CSA protons is small compared to TNSA, even under ideal conditions\cite{Haberberger2012}.
Also, TNSA expansion on the backside of the target spreads the shock accelerated ion's energy distribution such that they are indistinguishable from TNSA-only protons.

\begin{figure}[tb]
    \includegraphics[width=\columnwidth]{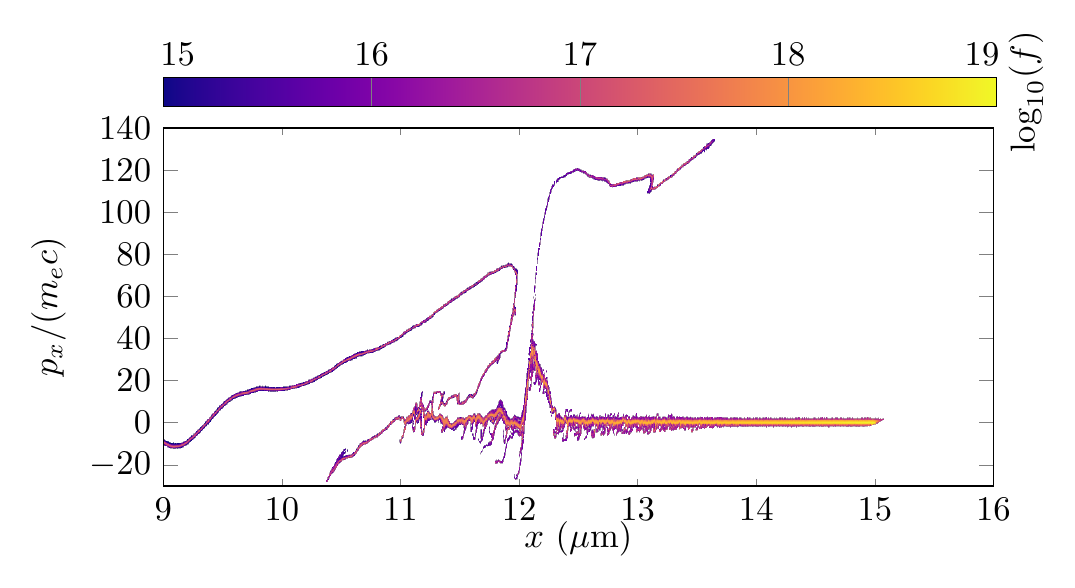}
    \caption{\label{fig:fsprot} The proton distribution on the frontside of a $12\unit{\mu m}$ thick target with pre-plasma, focused on the shock region at time $t=t_0+72.9t_L$. A similar shock is also observable on the $3\unit{\mu m}$ target.}
\end{figure}

The current simulations reiterate these findings: a shock front forms on the front side of the target, which can be seen in Figure \ref{fig:fsprot} as a bifurcation of the frontside proton distribution at around $12\unit{\mu m}$.
Figures \ref{fig:elecfield} and \ref{fig:elecfield12} illustrate the breakdown of this shock front which accelerates a small packet of frontside protons through the target.
However, this mechanism has little effect on the overall result.
The largest contribution of the CSA protons for the $12 \unit{\mu m}$ target is at $t = t_0 + 327t_L$, where a peak between 1.85 and 1.95 MeV hits $2.15\ten{14}$ MeV$^{-1}$ with an entire range between 1.8 and 3.1 MeV, generally reaching a count of $0.2\ten{14}$ MeV$^{-1}$ (not shown).
This concentration is too low and energy range too narrow to explain the plateau region via CSA protons alone.

From these findings, we argue that the low number density and homogenisation of the CSA protons do not contribute to plateau formation in any meaningful manner.

\section{The mechanism of plateau formation}\label{sec:plateau}

\begin{figure}[tb]
    \includegraphics[width=\columnwidth]{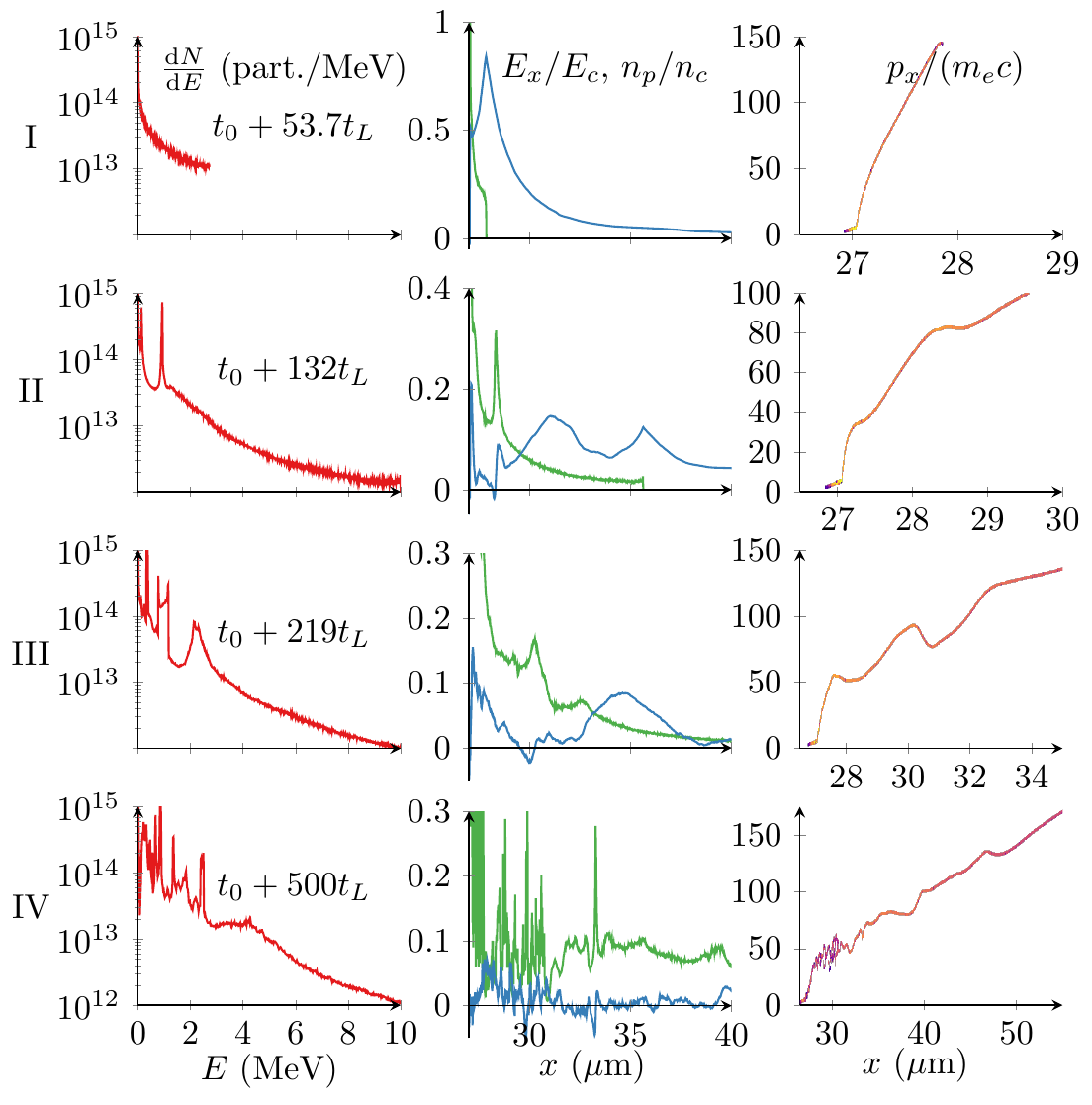}
    \caption{\label{fig:ionenergy} Energy peak spreading (left), electric field, backside proton density (center, blue and green respectively), as well as the dynamic regions of the proton $x$--$p_x$ phase space (right) for four time snapshots
        of the $12\unit{\mu m}$ target with a pre-plasma. See text for details. In the center frames, the electric field has been normalised here by the Compton field $E_c = m_e c \omega_L /e$. Whilst these frames show the electric field snapshots for each corresponding time slice, it's important to note that protons do not respond to the field instantaneously. As such there is a lag between the field and observed responses in the proton energy spectra. }
\end{figure}

We will now proceed to describe the process which, in our 1D simulations, causes energy plateaus in ion energy spectra.
Energy evolution of the $12\unit{\mu m}$ target and illustrations of the mechanism are displayed in Figure \ref{fig:ionenergy}.

Under stable TNSA, backside protons are accelerated via a potential due to the quasi-static sheath field, initially generating an energy spectra with very low energy and high probability (compared to later times).
If the field remains stable for a long enough time, we expect many of these protons to continue accelerating under a time-varying and non-uniform electric field, although many will interact in some form with the environment.

 Thus we expect to obtain a broad, disperse energy distribution of ions in this scenario, where only some of the accelerated particles reach maximum energies -- since the outermost ions observe the most intense portion of the sheath field (Row I in Figure \ref{fig:ionenergy}).

With the addition of a pre-plasma, fast electron generation by the laser pulse interaction is enhanced. These fast electrons observe a longer recirculation period, causing a drop in backside electron density if there are no more fast electrons being generated (which depends on the pulse duration).
The time-oscillating character of the electron charge density at the backside disrupts the sheath when only a low population of electrons are located there, leading to a temporary halt in the acceleration of backside protons.
The disruption is initially localised at the target backside, so protons with sufficient energy are still accelerating under what to them is still a stable field (Row II).
Soon after, the sheath recovers as electrons re-circulate and obtain a positive velocity.
At this point, more backside protons are swept up and begin to accelerate as expected.
The depleted zone in the sheath continues to propagate, not entirely in phase with the higher energy ions.
As the zone interacts with the peak, lower energy ions are not accelerated further (or perhaps even decelerated) at the same time as the higher energy ions continue to move under the still stable sheath (Row III) -- effectively spreading the energy peak into a plateau structure (Row IV).

\begin{figure}[tb]
    \includegraphics[width=\columnwidth]{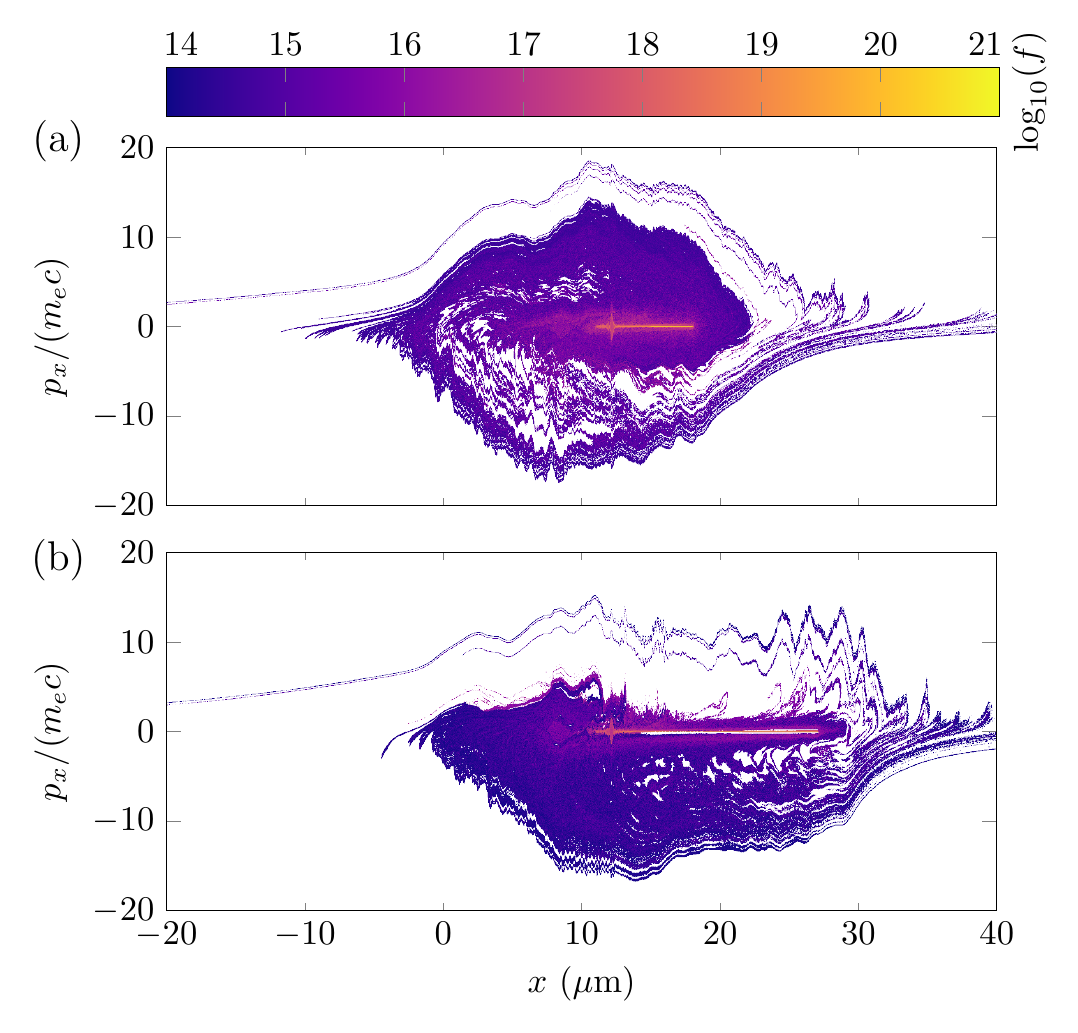}
    \caption{\label{fig:xpxcmp} Electron distributions of targets a)
      $3\unit{\mu m}$ and b) $12\unit{\mu m}$ thick with a
      pre-plasma. Time $t=t_0+72.4t_L$, lies in the middle of the
      disruption of the sheath field for the $12\unit{\mu m}$
      case. Compare the symmetry of each process and note that almost
      the entire fast electron population is travelling back toward the
      $12\unit{\mu m}$ target at this point in time. }
\end{figure}

This mechanism occurs in both the $3$ (thin) and $12 \unit{\mu m}$ (thick) targets, with Rows I and II of Figure \ref{fig:ionenergy} manifesting in exactly the same manner if we compare the distribution functions in Figures \ref{fig:flatcmp}b and \ref{fig:elecdist12}a.
The time under disruption is the discerning factor which answers most of the open questions about plateau formation.
Figure \ref{fig:xpxcmp} are electron distribution functions for the a) thin and b) thick targets at $t = t_0 + 72.4t_L$: in the middle of the first disruption phase of the thick target (see Figure \ref{fig:elecfield12}).

We see here that although the thin target has regions of low density in its electron distribution, the recirculation time for the electrons is much shorter allowing the majority of the fast electrons to mix and homogenise with the fast electrons still being generated on the front side by the tail end of the laser pulse interaction, weakening the TNSA field oscillations.
Electrons in the thick target however have only just begun to reach the frontside again and the laser pulse is no longer injecting new fast electrons into the distribution, yielding a distinct asymmetry in velocity space, generating a density perturbation and potential mitigating the acceleration of TNSA.
This portion of the process is Rows III and IV of Figure \ref{fig:ionenergy} which, in comparison to the case of the thin target, notably halts low energy proton acceleration and widens the initial peak to a conspicuous plateau region.
Since the TNSA process is almost completely suppressed for these $30$ laser periods (see Figures \ref{fig:elecdist12} and \ref{fig:elecfield12}), protons are under the influence of an intermittent and weakened accelerating potential compared to those in the thin target, which explains why higher maximum proton energies are observed experimentally in energy profiles without a plateau.

\begin{figure}[tb]
    \includegraphics[width=\columnwidth]{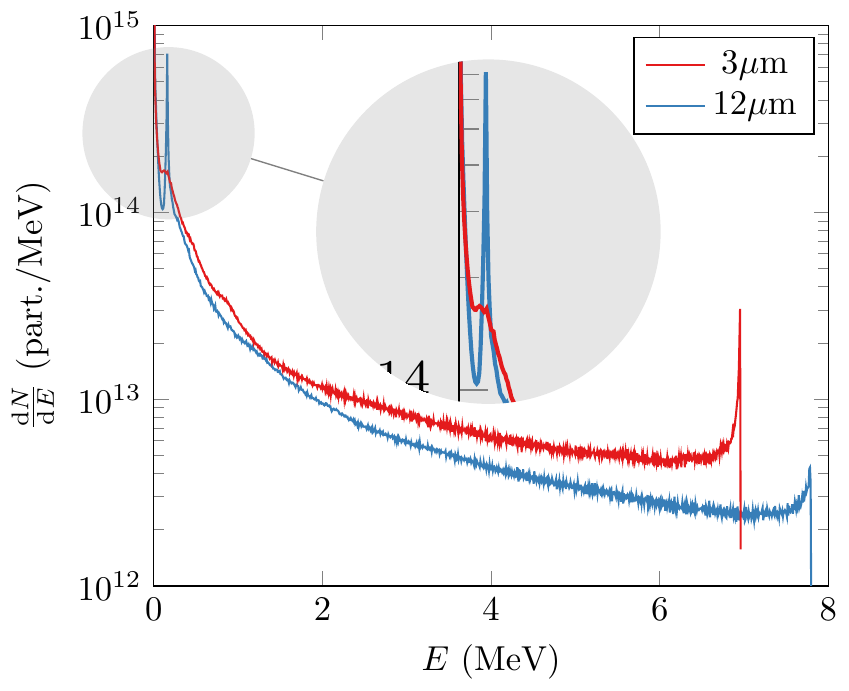}
    \caption{\label{fig:energy} Energy spectra comparison of two
      target thicknesses after a completed disruption even. The $3
      \unit{\mu m}$ target (red) regains a sheath field near
      $t=t_0+57.4t_L$ (see Figure \ref{fig:elecfield}) after a minimal
      disruption. This causes no obvious peaks in the spectrum and
      hence no plateau region develops in the final spectrum. The $12
      \unit{\mu m}$ target (blue) however, concludes a long disruption
      cycle near $t=t_0+94.9t_L$ (see Figure \ref{fig:elecfield12})
      and as a consequence develops a large energy peak that will
      develop into an energy plateau in its final spectrum. The peaks
      at high energy here are due to heavy-light ion interactions (see
      Section \ref{sec:species}).}
\end{figure}

Figure \ref{fig:energy} shows the energy spectra for both the thin and thick targets just after the initial disruption cycle has completed, showing a large spectral peak in the $12 \unit{\mu m}$ case.  This early peak will later evolve into an energy plateau, to the contrary of the regular energy spectra shown for the $3 \unit{\mu m}$ target.
Note that the $2$ MeV peaks discussed in relation to Figure \ref{fig:nrgad} occur relatively late in time, long after the time scale of this mechanism, therefore spreading of these peaks is not expected.

\begin{figure}[tb]
    \includegraphics[width=\columnwidth]{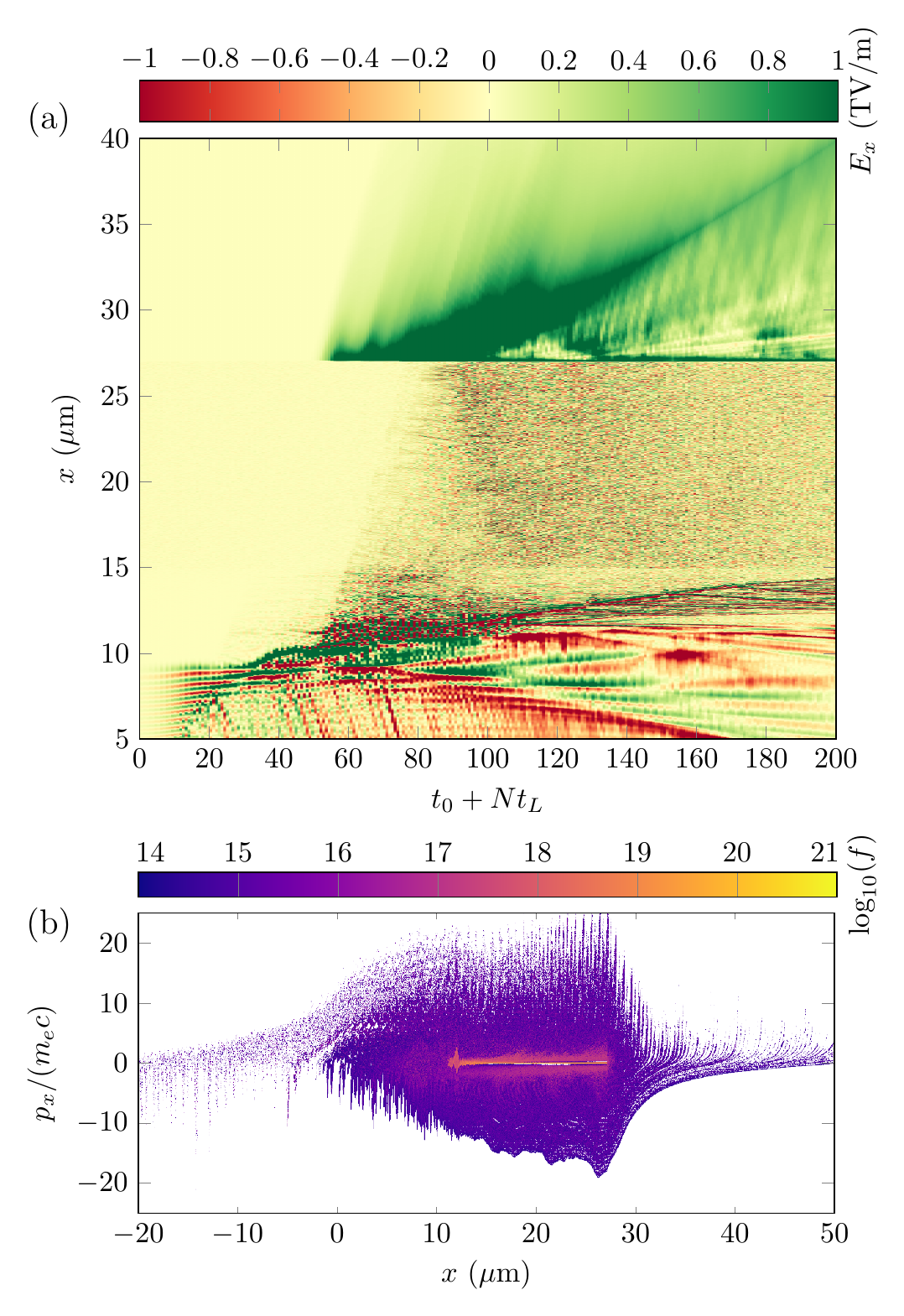}
    \caption{\label{fig:widepulse} a) Electric field in the $x$ direction
        for a $12\unit{\mu m}$ target with a pre-plasma and a $100$ fs pulse (FWHM). On the
        horizontal axis, $N$ represents the axis value of the time formula. Notice no sheath depletion due to recirculation occurs, just dissipation as the electrons escape and cool.
      b) Electron distribution function at $t=t_0+94.4t_L$ which, after rescaling to the sheath initialisation time, should be in the middle of a disruption event. Compare with Figure \ref{fig:xpxcmp}b and notice the symmetry disparity.}
\end{figure}

Control of the density depletion is possible by altering the ratio of the pulse time to the period of the fast electron current.
For example, using a pulse length of $t_{\mathrm{pulse}} = 100~$fs and a $12 \unit{\mu m}$ target should not observe a disruption (recall from Section \ref{sec:thickness} that $t_{\mathrm{sh}}\sim120$ fs for the $12\unit{\mu m}$ target) and therefore no plateau region should be observable in the energy spectrum.
Figure \ref{fig:widepulse}a displays the electric field of this case, where we see no attempt to re-establish a sheath after a disruption event, just the anticipated TNSA field peak and subsequent plateau in the quasineutral expanding plasma\cite{Mora2003} as the interaction evolves and target cools.
As the pulse duration is longer, initialisation of the sheath occurs at a later time than the $t_{\mathrm{pulse}}= 38$ fs case, although $t_{\mathrm{sh}}$ is not dependent on the pulse time.
Thus, if we observe the electron distribution within the disruption cycle (compare Figure \ref{fig:xpxcmp}), offset to the sheath initialisation time, we see the dynamics of Figure \ref{fig:widepulse}b.
The distribution is mostly symmetric in $p_x$ and causes fewer charge modulations since there is a continuous population density of electrons on the target backside.
Altering the pulse duration and verifying the absence of an energy plateau is therefore an experimentally verifiable signature to confirm this paper's hypothesis.

\section{Conclusions}

Laser setups where the field strength is $2 \lesssim a_0 \lesssim 10$ and contrasts in the range of (approximately) $10^{8}$ to $10^{10}$, shooting on thin targets with thicknesses of order $10 \unit{\mu m}$ observe a plateau in their ion energy spectra under the TNSA regime.
We identify a mechanism which explains this phenomenon that is dependent on pre-plasma interaction generating fast electrons from the low density region of the frontside of the target.
Temporal oscillations of the fast electron charge density at the backside of the target caused by electron recirculation disrupt the TNSA field for a short period causing peaks in the ion spectra, which are then spread in energy from a modified potential (due to electron recirculation) -- forming a plateau.
Disruption of the sheath field hinders continuous proton acceleration, therefore maximum proton energies are inhibited in comparison to experiments where no disruptions occur.
Protons accelerated by a shock on the frontside are only a minimal contribution to the final spectrum and are mostly irrelevant to the process.
However, the collapsing shock front may assist in further homogenisation and peak spreading as it dissipates its final energy.

The study has shown that pre-plasma interactions can play a non-trivial role in laser--matter experiments, and using clean slabs in simulations will definitely overlook many of these interactions.
An experimental signature connecting laser pulse and target widths has been identified, suggesting a way to control the plateau formation mechanism in the lab via alteration of the pulse duration to match a complete electron recirculation period.

We note that under realistic multidimensional conditions (i.e. with a finite laser focal spot), the intrinsic angular divergence of the hot electrons will cause them to expand transversely through the target, and thus become more and more diluted around the laser axis, where the fastest ions are driven \cite{Arefiev2016}.
In a target thin enough ($d < ct_\mathrm{pulse}/2$) that the sheath electric field does not undergo disruption events, this will mainly speed up the decay of the on-axis sheath field, reducing the maximum achievable proton energy.
In the thick-target case ($d > ct_\mathrm{pulse}/2$), which gives rise to periodic disruptions of the TNSA field, only a fraction of the hot electrons will return on axis after each recirculation period.
As a result, the disrupted sheath field will only partially recover, yielding weaker modulations in the resulting proton spectrum.
These dynamics are therefore expected to drastically diminish additional peak formation after the first disruption event, resulting in smoother proton energy spectra in the low energy limit (removing some of the peaks visible in energy spectra of Figures \ref{fig:nrgad} and \ref{fig:ionenergy}).

Furthermore, it is well-known that the efficiency and properties of hot-electron generation in undercritical preplasma may be sensitive to multidimensional effects.
An important mechanism to consider is the relativistic self-focusing of the laser pulse, leading  to an enhancement of its local intensity, and the radial expulsion of the plasma electrons by the laser ponderomotive force.
The electron dynamics in the resulting ion channel would then differ from that occurring in a 1D plasma, notably because of a possible coupling between the laser field and self-generated transverse (electron and magnetic) quasi-static fields, allowing for electron acceleration to super-ponderomotive energies \cite{Pukhov1999, Arefiev2016b}.
The scenario addressed in our paper, however, only relies on the production of an electron bunch energetic enough to drive efficient TNSA, regardless of its detailed energy spectrum.
The conclusions of the present work should therefore not be qualitatively impacted in a multidimensional setup.

Understanding the plateau formation process as a pre-plasma interaction and not an equipment misalignment will be helpful in the data analysis of current experiments.
Future experiment designs will also benefit from this information in a multitude of ways.
For example, high contrast campaigns (where pre-plasma is minimal) shooting on multi-layered targets comprised of a low density (under-dense) front side are expected to experience similar behaviour.

\begin{acknowledgments}
This work was supported by the Knut and Alice Wallenberg Foundation and the European Research Council (ERC-2014-CoG grant 647121).
 The simulations were performed on resources at Chalmers Centre for Computational Science and Engineering (C3SE) provided by the Swedish National Infrastructure for Computing (SNIC).
EPOCH was developed under UK EPSRC grants EP/G054950/1, EP/G056803/1, EP/G055165/1 and EP/M022463/1.
\end{acknowledgments}

\bibliography{plateau.bib}

\end{document}